\documentstyle[12pt,epsfig,epsf]{article}

\newlength{\dinwidth}
\newlength{\dinmargin}
\begin{document}

\def\bold#1{\setbox0=\hbox{$#1$}%
     \kern-.025em\copy0\kern-\wd0
     \kern.05em\copy0\kern-\wd0
     \kern-.025em\raise.0433em\box0 }
\def\slash#1{\setbox0=\hbox{$#1$}#1\hskip-\wd0\dimen0=5pt\advance
       \dimen0 by-\ht0\advance\dimen0 by\dp0\lower0.5\dimen0\hbox
         to\wd0{\hss\sl/\/\hss}}
\newcommand{\bea}{\begin{eqnarray}}
\newcommand{\eea}{\end{eqnarray}}
\newcommand{\be}{\begin{equation}}
\newcommand{\ee}{\end{equation}}
\newcommand{\nn}{\nonumber}
\newcommand{\dd}{\displaystyle}
\newcommand{\bra}[1]{\left\langle #1 \right|}
\newcommand{\ket}[1]{\left| #1 \right\rangle}
\newcommand{\spur}[1]{\not\! #1 \,}
\thispagestyle{empty}
\vspace*{1cm}
\rightline{BARI-TH/99-333}
\vspace*{2cm}
\begin{center}
  \begin{Large}
{\bf Radiative Leptonic $B_c$ Decays}\\
  \end{Large}
  \vspace{8mm}
  \begin{large}
P. Colangelo  \footnote{E-mail address: COLANGELO@BARI.INFN.IT},  
F. De Fazio \footnote{E-mail address: DEFAZIO@BARI.INFN.IT} \\
  \end{large}
  \vspace{6mm}
{\it Istituto Nazionale di Fisica Nucleare, Sezione di Bari, Italy}\\
\end{center}
\begin{quotation}
\vspace*{1.5cm}
\begin{center}
  \begin{bf}
  Abstract\\
  \end{bf}
\end{center}
\noindent
We analyze the radiative leptonic $B_c$ decay mode:
 $B_c \to \ell \nu \gamma$ ($\ell=e, \mu$) using
a QCD-inspired constituent quark model.
The prediction: 
${\cal B}(B_c \to \ell \nu \gamma)\simeq 3 \times 10^{-5}$ confirms that
this channel is experimentally 
promising in view of the large number of $B_c$
mesons which are expected to be produced at the future hadron facilities.

\vspace*{0.5cm}
\end{quotation}

\newpage
\baselineskip=18pt
\setcounter{page}{1}

Recently, the CDF Collaboration at the Fermilab Tevatron
has reported the  observation of the $B_c$ meson, 
the lowest mass ${\bar b}c$ ($b {\bar c}$) bound state, 
through the semileptonic  decay mode $B_c^\pm \to J/\psi \ell^\pm X$
\cite{cdf}. The  measured mass and lifetime of the meson are
\bea
M_{B_c}&=& 6.40 \pm 0.39\; ({\rm stat})\;
\pm  0.13 \; ({\rm syst})\; {\rm GeV} \label{mass}
\\
\tau(B_c) &=& 0.46^{+0.18}_{-0.16} \; ({\rm stat}) 
\pm 0.03 \; ({\rm syst}) \; {\rm ps}
 \;\;\;. \label{tau} 
\eea
The particular interest of this observation is related to the 
fact that the meson ground-state with open beauty and charm can decay 
only weakly, thus
providing the rather unique opportunity of investigating weak decays
in  a heavy quarkonium-like system. 
Moreover, 
studying this meson, important information can
be obtained,  not only concerning fundamental
parameters, such as, for example,
 the element $V_{cb}$ of the Cabibbo-Kobayashi-Maskawa
mixing matrix, but also the strong dynamics responsible of the
binding of the quarks inside the hadron.
Understanding such dynamics is one of the most important
issues in the  analysis of heavy hadrons \cite{gershtein}.
The observation of the $B_c$ meson
at the Tevatron confirms that  $B_c$ physics will gain an important role 
at the future hadron facilities, where a large production rate of such
particles
is expected; in particular, at the Large Hadron Collider (LHC), which will
be operating at CERN, it is estimated that
$4.5 \times 10^{10}$ $B_c^+$ mesons 
will be produced per year for a machine luminosity of 
${\cal L}=10^{34}$ cm$^{-2}$ sec$^{-1}$ at ${\sqrt s}= 14$ TeV
\cite{cheung}.

$B_c$ meson decays can be classified according to the mechanism inducing the
processes at quark level. Neglecting Cabibbo-suppressed and
penguin-induced transitions, such mechanisms are:
\begin{itemize}
\item the  $b$-quark transition 
$b \to c \; W^-$,  with the ${\bar c}$ quark having
the role of spectator; the corresponding final states are of the kind 
$J/\psi \; \pi$, $J/\psi \; \ell \nu$;
\item the charm quark transition 
${\bar c } \to {\bar s} \; W^-$, with $b$ as spectator  and
possible final states  $B_s \; \pi$, $B_s \; \ell \nu$, etc.;
\item the annihilation modes ${\bar c } b \to W^-$.
\end{itemize}

The first two mechanisms are responsible of the largest part of the $B_c$
decay width \cite{lus,quigg,beneke}. In particular,
the measurement  (\ref{tau}) provides us with an indication that
the dominant $B_c$ decay mechanism is the $c$-quark decay, which implies
a $B_c$ lifetime in the range $\tau_{B_c}=(0.4-0.7)\;ps$ \cite{beneke}, 
whereas dominance of the $b-$quark decay mechanism would produce
a longer lifetime: $\tau_{B_c}=(1.1-1.2)\;ps$  \cite{quigg}.  
Various analyses of $B_c$ 
transitions induced by the two mechanisms are available in the literature
\cite{gershtein}; for example,
a QCD sum rule calculation of  $B_c$ semileptonic decays
suggested the dominance of the charm transition \cite{paver}.

As far as  the annihilation processes are concerned, 
the leptonic radiative decay mode $B_c \to \ell  \nu \gamma$ 
and  the leptonic decay without photon in the final state
represent  a minor fraction of the $B_c$ full width.  Nevertheless,
their analysis is of particular interest, both from the  
phenomenological and the theoretical point of view.
From the phenomenological side, 
$B_c$ annihilation modes are governed by $V_{cb}$;  therefore they are
Cabibbo-enhanced with respect to the analogous $B_u$ decays, and represent
new channels  to access this matrix element. 
From the theoretical viewpoint, the  purely leptonic 
and the radiative leptonic $B_c$ transitions
are interesting since, in the nonrelativistic limit of the 
quark dynamics, both their rates can  be expressed in terms of a single
hadronic parameter, the $B_c$  leptonic decay constant $f_{B_c}$
\cite{cinesi,falk}. In this limit, 
a  relation between the widths of the processes
$B_c \to \ell  \nu \gamma$ and $B_c \to \ell  \nu$ 
can be worked out \cite{cinesi,falk}:
\be
R_\ell={\Gamma (B_c \to \ell \nu \gamma) \over \Gamma (B_c \to \ell \nu)}
\simeq 0.40 \; r_\ell \label{ratio}
\ee
with $\dd{r_\ell={\alpha \over 4\pi} {m_{B_c}^2 \over m_\ell^2}}$.
Eq.(\ref{ratio}) implies that
 the width of the radiative leptonic $B_c$  decay into muons 
is nearly equal to the purely leptonic
width  $\Gamma(B_c \to \mu \nu)$, whereas, in the case of
electrons in the final state, the radiative leptonic mode is largely dominant.

Eq.(\ref{ratio})
presents uncertainties coming from the used values of the charm and beauty 
quark masses. Moreover, there
could be  corrections if the quark 
dynamics in the $B_c$ meson 
deviates from the nonrelativistic regime, and the size of the corrections
is useful for understanding  the theoretical uncertainty affecting the ratio
(\ref{ratio}).

Corrections to the ratio (\ref{ratio}) can be estimated
by considering a model
for the $B_c$ meson where relativistic effects 
in the constituent quark dynamics are, at least partially, taken into account; 
the present letter is devoted to such a study.

In order to analyze the decay mode $B_c \to \ell \nu \gamma$ ($\ell=e, \mu$), 
we follow the method 
adopted in ref.\cite{noi} to investigate the  analogous  $B_u$
transition. Let us consider  the process
\begin{equation}
B_c^-(p) \to \ell^-(p_1) \; { \bar \nu}(p_2) \; 
\gamma(k, \epsilon) 
\label{channel}
\end{equation}
whose amplitude can  be written as
\be
{\cal A}(B_c \to \ell { \nu} \gamma)
={G_F \over \sqrt{2}} V_{cb} \;\;\epsilon^{* \nu} \; \; 
\big( L^\mu \cdot \Pi_{\mu\nu} -i e f_{B_c} p^\mu \tilde L_{\mu\nu} \big) \hskip
10 pt, \label{amp} 
\ee
with $G_F$  the Fermi constant
 and $\epsilon$  the photon polarization vector.
The current $L^\mu={\bar \ell}(p_1) \gamma^\mu (1-\gamma_5)\nu(p_2)$
represents the weak leptonic current in (\ref{channel}).  
The hadronic function $\Pi_{\mu \nu}(p,k)$ is the correlator
\be
\Pi_{\mu \nu}(p,k)= i\; \int d^4 x
e^{i q \cdot x} <0|T[ J_\mu(x) V_\nu(0) ]|B_c(p)>  \hskip 3 pt 
\label{pi}
\ee   
with $J_\mu(x)={\bar c}(x) \gamma_\mu (1 -\gamma_5) b(x)$
the weak hadronic current governing (\ref{channel}) and $V_\nu$ given by
$V_\nu(0)=e [ Q_c \; {\bar c}(0) \gamma_\nu c(0)+
Q_b \; {\bar b}(0) \gamma_\nu b(0)]$, $e Q_c$ and $e Q_b$ being the 
charm and beauty  quark electric charges.
The momentum $q$ is defined as $q=p_1+p_2$. 
Therefore, the first term in (\ref{amp}) correspond to the photon 
emitted from the 
meson, and the second term  represents the contribution of the photon
emitted by the charged lepton leg.
Notice that the $B_c$ leptonic constant $f_{B_c}$ is
defined by the matrix element
\be
<0|\bar c \gamma_\mu \gamma_5 b| B_c(p)> = i f_{B_c} p_\mu 
\label{fB}
\ee
and that $\tilde L_{\mu\nu}$ reads
\be
\tilde L_{\mu\nu}= 
{\bar \ell}(p_1) \gamma^\nu 
{1 \over {\spur p}_1 + {\spur k} -m_l} \gamma^\mu (1-\gamma_5)\nu(p_2) \;\;.
\ee

The hadronic function  $\Pi_{\mu \nu}(p,k)$ can be expanded in independent 
Lorentz structures
\be
\Pi_{\mu \nu}(p \cdot k) = 
\alpha \; p_\mu p_\nu+
\beta \; k_\mu k_\nu+
\zeta \; k_\mu p_\nu+
\delta \; p_\mu k_\nu+
\xi \; g_{\mu \nu} +
i \eta \epsilon_{\mu \nu\rho\sigma}p^\rho k^\sigma \;\,\;
\label{invariants}
\ee
(with the invariant functions $\alpha, \dots \eta$ depending on $p \cdot k$)
so that the requirement of gauge invariance for the amplitude (\ref{amp})
implies the condition:
\be
(\alpha+\zeta)  (p \cdot k) +\xi- i e f_{B_c} = 0 \;\;\; .
\label{gauge}
\ee
We shall see in the following that this condition is satisfied 
in our calculation.

The final expression of the amplitude, eq.(\ref{amp}),
\be
{\cal A}(B_c \to \ell { \nu} \gamma)
={G_F \over \sqrt{2}} V_{cb} \;\;\epsilon^{* \nu} \; L^\mu
[ (\alpha+\zeta)\;
( k_\mu p_\nu - p \cdot k \; g_{\mu \nu}) + i \; \eta
 \; \epsilon_{\mu \nu \rho \sigma} p^\rho k^\sigma ] \; \epsilon^{* \nu} 
\;\;\; , \label{amp1}
\ee
is given in terms of the form factors  $\eta$ and 
$\zeta+\alpha$, related to the
vector and the axial vector  weak current contributions to the process
(\ref{channel}), respectively.

In order to compute  the invariant functions in (\ref{amp1}) we 
employ the costituent quark model developed in 
ref.\cite{pietroni} to describe the static
properties of mesons containing heavy quarks. 
In the correlator (\ref{pi}) we write the state of the pseudoscalar
$(b {\bar c})$ meson at rest in terms of a wave-function $\psi_{B_c}$
and of quark and antiquark creation operators:
\be |B_c^->=i {\delta_{\alpha \beta} \over \sqrt{3}}
{\delta_{rs} \over \sqrt{2}}
\int d \vec{k} \; \psi_{B_c} (\vec{k})\;
b^{\dag}(\vec{k}, r, \alpha) \;
c^{\dag}(-\vec{k}, s, \beta)|0> \hskip 5 pt ; \label{b}
\ee
$\alpha$ and $\beta$ are colour indices, $r$ and $s$ spin indices;
the operator $b^{\dag}$ creates a $b$ quark  with momentum
${\vec k} $, while $c^{\dag}$ creates a charm antiquark
with momentum $-\vec{k}$.
The wave-function $\psi_{B_c}$, 
describing the quark momentum distribution in the meson,
is obtained as a solution of the
Salpeter equation 
\be
\Big\{ \sqrt{\vec{k}^2 + m_b^2} + \sqrt{\vec{k}^2 + m_c^2}-m_{B_c} \Big\}  
\psi_{B_c} ( \vec{k})
+ \int d \vec {k^{\prime}} \; V(\vec{k}, \vec{k^{\prime}}) \;
  \psi_{B_c} (\vec{k^{\prime}})=0 
\label{7} 
\ee
stemming from the quark-antiquark
Bethe-Salpeter equation, in the approximation of an
istantaneous interaction represented by the interquark potential $V$.  
Within the model in \cite{pietroni}, 
$V$ is chosen as the Richardson potential
 \cite{richardson}, which reads in the $r-$space:
\be
V(r)={8 \pi \over 33 -2 n_f} \Lambda \Big[ \Lambda r-{f(\Lambda r) \over
\Lambda r} \Big] \;\;\; , \label{pot}
\ee
\noindent with $\Lambda$  a parameter,
$n_f$  the number of active flavours, and the function $f(t)$  given
by:
\be
f(t)={4 \over \pi} \int_0^\infty dq {sin(qt) \over q} \Big[ {1 \over
ln(1+q^2)}
- {1 \over q^2} \Big] \; . \label{f}
\ee
\noindent
The linear increase at large distances of the potential in (\ref{pot})  
provides QCD confinement; at short distances the potential
behaves as $- {\alpha_s(r) \over r}$,
with $\alpha_s(r)$ logarithmically decreasing with the
distance $r$, thus reproducing the asymptotic freedom property of QCD.
A smearing procedure at short-distances is also adopted,
to account for effects
of the relativistic kinematics \cite{pietroni}. Finally,
spin interaction effects are neglected since
in the case of heavy mesons the chromomagnetic coupling is of order
of the inverse heavy quark masses.

The interest for  eq.(\ref{7}) is that 
relativistic effects are taken into account at least in the quark kinematics. 
Therefore, the same wave-equation  can be used to study heavy-light
as well as heavy-heavy quark systems,  and one case can be obtained 
from the other one by continuously varying a parameter, the 
$u(c)$ quark mass. From the analysis of the solutions,
a comparison between the two cases, the heavy-light and the heavy-heavy
quark meson systems,
can be meaningfully performed.

Eq.(\ref{7})  can  be solved by numerical methods, as described in
\cite{pietroni}, fixing 
the masses $m_c$ and $m_b$ of the constituent quarks,
together with  the parameter $\Lambda$, in such a
way that the charmonium and  bottomomium spectra 
are reproduced. The chosen values  for the parameters are: $m_b= 4.89$ 
GeV and  $m_c=1.452$ GeV, with $\Lambda=397$ MeV \cite{pietroni}.
A fit of the heavy-light meson masses also fixes the
values of the constituent light-quark masses: $m_u=m_d=38$  MeV
\cite{pietroni}. 
For the $b \bar c$ system all the input parameters required in 
(\ref{7}) are fixed from the analysis of other channels, and the predictions 
do not depend on new external quantities. The numerical solution of (\ref{7})
produces  spectrum of the $b \bar c$ bound states; 
the predicted masses of the 
first three radial $S-$wave resonances are reported in Table 1.
\begin{table}[ht]
\begin{center}
\baselineskip=12pt
\vskip 0.3 cm
\baselineskip=18pt
\begin{tabular}{|c|c|c|}
\hline \hline
radial number & mass (GeV) & leptonic constant (MeV) \\ \hline \hline
$0$ & 6.28    & 432  \\ \hline
$1$ & 6.80    & 313  \\ \hline
$2$ & 7.16    & 265  \\ \hline
\end{tabular}
\end{center}
\label{table1}
\caption{Spectrum and leptonic constants of the pseudoscalar $b \bar c$ 
mesons.}
\end{table}

The spectrum in Table 1 agrees with  other theoretical determinations 
based on constituent quark models
\cite{qmodels}. As for QCD sum rule and lattice QCD results, in
\cite{paver} the value $m_{B_c}=6.35$ GeV was found using two-point
function QCD sum rules, whereas 
a recent lattice QCD  analysis predicts:
$m_{B_c}=6.388\pm 9\pm 98\pm 15$ GeV, with the larger error related  to the 
quenched approximation \cite{shanahan}. Within the errors, the mass of the
$B_c$ meson  in Table 1 agrees with the CDF 
 measurement reported in eq.(\ref{mass}).

Also the wave-function $\psi_{B_c}$ can be obtained by solving eq.(\ref{7}).
We use the  covariant normalization:
\be {1 \over (2\pi)^3} \int d \vec{k} |\psi_{B_c}(\vec k)|^2=
2 m_{B_c} \hskip 3 pt  \label{eq : 8} \ee
\noindent
and define, in the $B_c$ meson rest-frame, 
the reduced wave-function $u_{B_c}(k)$ ($k=|{\vec k}|$):
\be u_{B_c}(k)={k \; \psi_{B_c}(k) \over \sqrt{2} \pi } \hskip 3 pt
\label{eq : 18} \ee
\noindent which  is normalized as: $\int_0^\infty dk |u_{B_c}(k)|^2=2
m_{B_c}$.
In fig.1 the $B_c$ meson wave-function $u_{B_c}(k)$ is depicted, 
together with the function $u_{B_u}$ of the $B_u$ meson.
\begin{figure}[ht]
\begin{center}
\vskip -2cm
\epsfig{file=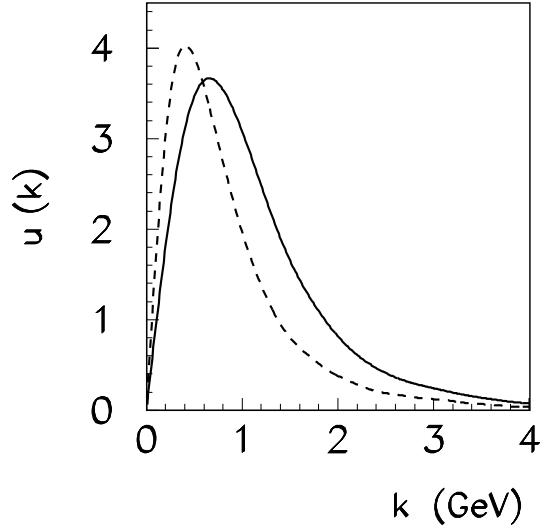,width=13cm} 
\label{wavefunc}
\vskip -3.0cm
\caption{$B_c$ meson wave-function (continuous line). 
The wave-function of the $B_u$ meson is also depicted (dashed line).}
\end{center}
\end{figure}

The meson wave-functions, together with the predictions of the spectrum of the
bound states, are the main results of the model; they allow us to calculate
hadronic quantities such as the
leptonic decay constants and the strong couplings to light mesons
\cite{pietroni,defazio94}. In particular, using 
$u_{B_c}(k)$ depicted in fig.1, we can infer the size of the deviation from the
nonrelativistic limit in the Salpeter equation (\ref{7}). As a matter of fact,
the average squared quark momentum $<k^2>$
turns out to be $<k^2>=0.95$ GeV$^2$, and the ratios  $<k^2>/m_c^2=0.43$ and
$<k^2>/m_b^2=0.04$. In the case of the $B_u$ meson,
the average squared quark momentum  $<k^2>$ is  $<k^2>=0.47$ GeV$^2$, to be 
compared to $m_u^2=1.4\times 10^{-3}$ GeV$^2$. Therefore, in the case of $B_c$
mesons, deviations from the nonrelativistic limit, although small, are not
negligible, mainly due to term related to the charm quark. 

Before coming  to the analysis of the radiative leptonic $B_c$  decay 
and to the calculation of the ratio (\ref{ratio}),
let us evaluate  the leptonic constants $f_{B_c^n}$ of the 
$S$-wave excitations,  defined  by matrix elements analogous to (\ref{fB}).
The numerical results, obtained from  the eigenfunctions of eq.(\ref{7}),
are  reported in Table 1.
Few comments are in order. The value of $f_{B_c}$ in Table 1 is compatible,
within the theoretical errors, with the result $f_{B_c}=360 \pm 60$ MeV
obtained from QCD sum rules \cite{paver} (the uncertainty related to
the input parameters of the potential model is estimated of $\pm 15 \%$ 
for the leptonic constants \cite{pietroni}). On the other hand, the leptonic
constants in Table 1 are  smaller than the outcome of several
constituent quark models considered in \cite{qmodels}, 
based on a purely nonrelativistic description of the $b \bar c$ system.
Notice that the decreasing values 
of $f_{B_c^n}$, for increasing  radial number $n$,
 is due to the nodes of the wave-functions of the radial excitations. Finally, 
the $B_c$ leptonic constant turns out to be compatible
with a lattice NRQCD determination  \cite{woloshyn}.

Let us now consider the correlator (\ref{pi}).
Writing  the state $|B_c>$  as in  (\ref{b}) and
expanding the T-product according to the Wick theorem, we can express the
product of the two currents $J_\mu(x) V_\nu(0)$ in terms of  
quark creation and annihilation operators; then,  exploiting the
anticommutation relations between such operators, we  can derive
 an expression for
$\Pi_{\mu\nu}$  in terms of the $B_c$ wave function, $u_{B_c}$, which is analogous to the expression reported in \cite{noi} for the $B_u$ decay.

It is useful to relate the invariant functions in (\ref{invariants})
to the various components of the hadronic tenson $\Pi_{\mu\nu}$.
In the $B_c$ rest-frame
$p=(m_{B_c}, {\vec 0})$ and choosing $k^\mu=(k^0,0,0,k^0)$ one gets
from eq.(\ref{pi}):
\bea
\zeta(k^0)&=&{1 \over  m_{B_c} k^0} (\Pi_{30}-\Pi_{33} + \Pi_{11})   
\nonumber \\
\alpha(k^0)&=&{1 \over  m_{B_c}^2} (\Pi_{00}+\Pi_{33} -\Pi_{30}-\Pi_{03}) 
   \nonumber \\
\eta(k^0) &=& -i { 1 \over  m_{B_c} k^0}  \Pi_{12} (k^0) \hskip 10 pt .
\label{par}
\eea
Therefore, the condition (\ref{gauge}) ensuring gauge invariance
can be written as
\be
i e f_{B_c}= {k^0 \over m_{B_c}} 
\big(\Pi_{00}+\Pi_{33} -\Pi_{30}-\Pi_{03}\big)
+ \Pi_{30}-\Pi_{33} \;\;\;, \label{gauge1}
\ee
a condition that must be checked in our analysis.
The explicit calculation, using
the  expression of $\Pi_{\mu\nu}$
in terms of the meson wave-function  \cite{noi}, shows that 
eq.(\ref{gauge1}) is verified provided that
the  leptonic constant $f_{B_c}$ is given by
\be
 f_{B_c} = \sqrt{3} {1 \over 2 \pi m_{B_c}} 
\int_0^\infty dk^\prime \; k^\prime \; u_{B_c}(k^\prime) 
\big[ {(E_b+m_b)(E_c+m_c)\over E_b E_c}\big]^{1\over 2}
\Big[1- {k^{\prime2}\over (E_b+m_b)(E_c+m_c)}\Big]
\label{ff}
\ee
with $E_{b,c}=\sqrt{k^{\prime 2}+m_{b,c}^2}$. Indeed, eq.(\ref{ff}) 
is the expression for
$f_{B_c}$ obtained in the framework of the
constituent quark model \cite{pietroni}, hence the gauge invariance property of
the amplitude (\ref{amp}) is preserved in our calculation.
\footnote{In the case of the radiative leptonic $B_u$ decay,
the contribution proportional to $f_B$ turns out to be numerically negligible.}

Having checked the property of gauge invariance, we can simply compute the 
decay width of the mode (\ref{channel}) and the photon energy distribution. 
Notice that we compute
the photon energy spectrum for a photon energy larger than 
$100$ MeV, 
which represents the photon energy resolution we assume in our analysis.

The expression  for the width of the decay
(\ref{channel}), considering massless leptons in the final state 
($m_\ell \simeq 0$),  is
\be
\Gamma(B_c \to \ell {\nu} \gamma)= {G_F^2 |V_{cb}|^2 \over
3 (2 \pi)^3} \int_0^{m_{B_c}/2} dk^0 k^0 (m_{B_c}-2k^0)
[|\Pi_{11}+i e f_{B_c}|^2 +
|\Pi_{12}|^2] \hskip 3 pt . \label{gamma}
\ee
Using the data in (\ref{mass}),(\ref{tau}), 
together with $V_{cb}=0.04$, we predict: 
\be
\Gamma(B_c \to \ell \nu \gamma) = 4 \;10^{-16} GeV \hskip 0.5cm , 
 \hskip 0.5cm {\cal B}(B_c \to \ell \nu \gamma) = 3 \;10^{-5} \;.
\label{result}
\ee
The energy spectrum of the photon,
computed using eq.(\ref{gamma}),
is depicted in fig.2; it looks  symmetric with respect to the point 
$E_\gamma=m_{B_c}/4$, as observed also in \cite{cinesi,falk,aliev}.
\begin{figure}[ht]
\begin{center}
\vskip -2cm
\epsfig{file=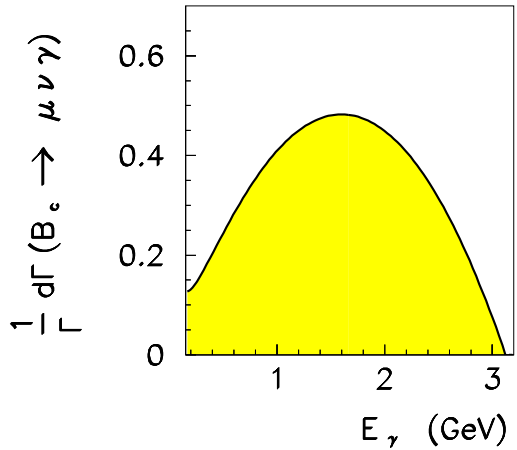,width=13cm} 
\label{specphoton}
\vskip -3.0cm
\caption{Photon energy spectrum in the decay (\ref{channel})}.
\end{center}
\end{figure}

The result (\ref{result}) confirms that the
rate of the radiative leptonic $B_c$ decays into muons (and electrons)
is sizeable, and could be accessed at the future hadronic machines, 
such as LHC.

Let us now consider  the ratio (\ref{ratio}). In order to determine it,
 we need the width of the
purely leptonic $B_c$ decay into electrons and muons, given by
\begin{equation}
\Gamma(B_c \to \ell \nu)={G_F^2 \over 8 \pi} |V_{cb}|^2
f_{B_c}^2
\Bigg( {m_\ell \over m_{B_c}} \Bigg)^2 m_{B_c}^3 \Bigg( 1-{m_\ell^2 \over
m_{B_c}^2}\Bigg)^2 \hskip 3 pt,  \label{width}
\end{equation}
which depends on $f_{B_c}$. Using the value  in Table 1 we obtain: 
\be
\Gamma(B_c \to \mu \nu)=1.1 \; 10^{-16} \; GeV \hskip 0.5cm , 
\hskip 0.5cm
{\cal B}(B_c \to \mu \nu)=7.8 \; 10^{-5} \;\;\;,
\ee
 corresponding to   
$\displaystyle 
R_\mu={\Gamma(B_c \to \mu \nu \gamma) \over \Gamma(B_c \to \mu \nu)}\simeq
0.4$.  This result must be compared  to the value
$R_\mu \simeq 0.8$ obtained using eq.(\ref{ratio})
\cite{cinesi,falk}, and suggests  
that the corrections to the
nonrelativistic limit, although small, 
play a  role in modifying the prediction 
(\ref{ratio}). In any case, the widths of the radiative leptonic and the purely
leptonic $B_c$ transitions are still comparable, a result which is very 
different from the case of radiative 
leptonic $B_u$ decays, which are enhanced with respect to 
the purely
leptonic decay modes into muons and electrons. In that latter case, 
the  helicity suppression, displayed  by the factor 
$\big( {m_\ell \over m_{B_q}} \big)^2$ in (\ref{width}) and  producing 
very small decay rates, is efficiently
avoided by the presence of the photon in the 
final state also in the case of muons, so that
the branching fraction of the radiative leptonic $B_u$ modes into muons
turns out to be  enhanced  by  one order of magnitude
with respect to the purely muon leptonic decay \cite{noi,bmunugamma}.

As for the decay into electrons,  we predict
${\cal B}(B_c \to e \nu_e ) \simeq 2 \times 10^{-9}$.

The decay mode $B_c \to \ell \nu \gamma$
has been investigated, as mentioned above,  in the framework of the
nonrelativistic quark model (NRQM) \cite{cinesi,falk}, and also using
the light front quark model (LFQM) \cite{lih} and 
light-cone QCD sum rules (LCSR) \cite{aliev}. In
Table 2 we report the various predictions, obtained using the  value of
$f_{B_c}$ reported in Table 1.
We also report the estimates  of the ratio
\be
r={N_{B_u} \over N_{B_c}}={{\cal B}(B_u \to \ell \nu \gamma) \over
{\cal B}(B_c \to \ell \nu \gamma)} \label{ratiobis}
\ee
which represents the relative fraction of the final state $\ell \nu \gamma$
coming from the different sources $B_u$ and $B_c$.
\begin{table}[ht]
\begin{center}
\baselineskip=12pt
\vskip 0.3 cm
\baselineskip=18pt
\begin{tabular}{|c|c|c|c|}
\hline \hline
${\cal B}(B_c \to \ell \nu \gamma)$ &r & method &ref. \\ \hline \hline
$3  \times 10^{-5}$ &$0.03$   &Rel. QM        &This paper \\ \hline
$6\times 10^{-5}$ &          & NRQCD &\cite{falk} \\ \hline 
$6\times 10^{-5}$ &$0.06-0.1$& NRQM & \cite{cinesi} \\ \hline
$2\times 10^{-5}$ & & LFQM & \cite{lih} \\ \hline
$2\times 10^{-5}$ &$0.1$     & LCSR &\cite{aliev} \\ \hline
\end{tabular}
\end{center}
\label{table2}
\caption{Results for ${\cal B}(B_c \to \ell \nu \gamma)$ by different
approaches.}
\end{table}
Our small result: $r=0.03$  means that a  large number of 
$\ell \nu \gamma$ final states should come from $B_c$ decays;
the actual  fraction of radiative leptonic final states can be obtained
by multiplying eq.(\ref{ratiobis}) by the probabilities of producing
$B_u$ and $B_c$ mesons from $b$ quarks.

Let us conclude our analysis, based on a QCD-inspired relativistic
constituent quark model, observing 
that the quark dynamics inside the $B_c$ meson could  
 modify  the prediction (\ref{ratio}). Therefore,
the range $[0.4 - 0.8 ]$ for the ratio (\ref{ratio}) can be
interpreted as the theoretical uncertainty for this quantity. 
The measurement of the radiative leptonic $B_c$ decay
rate, together with the purely
leptonic rate, although challenging from the experimental viewpoint, 
is one of the expected results in the LHC era.

\vspace*{2cm}
\noindent {\bf Acknowledgments\\}
\noindent 
We thank G. Nardulli and N. Paver for discussions.

\end{document}